**Polar phonons in some compressively stressed epitaxial and polycrystalline $SrTiO_3$ thin films**

D. Nuzhnyy[1], J. Petzelt[1], S. Kamba[1], T. Yamada[2], M. Tyunina[3], A. K. Tagantsev[2], J. Levoska[3], N. Setter[2]

[1] *Institute of Physics ASCR, Na Slovance 2, 18221 Prague 8, Czech Republic, nuzhnyj@fzu.cz*

[2] *Ceramics Laboratory, Swiss Federal Institute of Technology, EPFL, CH-1015 Lausanne, Switzerland*

[3]*Microelectronics and Materials Physics Laboratories, and EMPART Research Group of Infotech Oulu, University of Oulu, PL 4500, FIN 900014 Oulun Yliopisto, Finland*

**Abstract:**

Several $SrTiO_3$ (STO) thin films without electrodes processed by pulsed laser deposition, of thicknesses down to 40 nm, were studied using infrared transmission and reflection spectroscopy. The complex dielectric responses of polar phonon modes, particularly ferroelectric soft mode, in the films were determined quantitatively. The compressed epitaxial STO films on (100) LSAT substrates show strongly stiffened phonon responses, whereas the soft mode in polycrystalline film on (0001) sapphire substrate shows a strong broadening due to grain boundaries and/or other inhomogeneities and defects. The stiffened soft mode is responsible for a much lower static permittivity in the plane of the compressed film than in the bulk samples.

$SrTiO_3$ (STO) is a classical incipient ferroelectric with strongly increasing permittivity on cooling – up to more than 20,000 in single crystals [1]. It is well known since 1994 [1] that the permittivity in STO thin films is strongly reduced and does not increase on cooling as much as in single crystals. It is thickness dependent, but even in the limit of very thick films it does not reach the crystal values [2]. The thickness dependence was assigned to low-permittivity (dead) interfacial layers between the film and electrodes [2]. However, grain boundaries also play an important role in bulk ceramics as well as in thin films [3-6]. Low-permittivity grain boundaries and possible nano-cracks in the films and/or ceramics decrease the effective permittivity, which is accompanied by a remarkable stiffening of the effective soft mode.

Strain induced by the substrate may also influence the dielectric response of epitaxial films. The phase diagram of STO under uniaxial (biaxial) pressure was theoretically calculated from the known thermodynamic parameters [7, 8]; it appears to be very sensitive to both compressive and tensile stresses. The ferroelectric transition may be also induced by the choice of appropriate substrate. For instance, the tensile in-plane stress produced by a (110) $DyScO_3$ substrate may induce the in-plane ferroelectricity up to room temperature [9]. Quasi-epitaxial film on (0001) sapphire exhibited ferroelectric transition near 120 K as evidenced by a softer $TO_1$ mode above 120 K compared to the single crystals [5]. Below the structural transition occurring probably at the same temperature, the $TO_1$ mode couples with the structural soft mode doublet and hardens on further cooling. On the other hand, the compressive stress induced by a (110) $NdGaO_3$ (NGO) substrate is responsible for the out-of plane ferroelectricity near 150 K [10] whereas the in-plane mode softening with decreasing temperature is only very weak. In the latter case, the in-plane $TO_1$ mode stiffens up to 132 $cm^{-1}$ at room temperature and softens only to 108 $cm^{-1}$ at low temperatures. The out-of-plane ferroelectric transition was indicated by the appearance of a weak silent $TO_3$ mode near 318 $cm^{-1}$ in the infrared (IR) spectra. [10].

In this paper we report the IR spectra of epitaxial STO films on (100) $La_{0.18}Sr_{0.82}Al_{0.59}Ta_{0.41}O_3$ (LSAT) single crystal substrates and of thin (~40 nm) polycrystalline STO film on a (0001) sapphire substrate, both grown by pulsed-laser deposition [10, 11]. The IR spectroscopy has proven to be a very powerful technique for studying the polar phonon properties, particularly soft modes, in thin ferroelectric films [5, 12]. Until recently, only transmission spectroscopy in the far IR was used, which limited the technique to the use of far-IR transparent substrates, such as sapphire, MgO, highly resistive Si, $SiO_2$ etc. without electrodes. For opaque substrates we have recently applied also a specular reflection technique to detect the polar phonons of thin STO films [10] and demonstrated that the technique can be used down to ~50 nm film thickness.

For this study, STO films were deposited on (100) LSAT substrate with KrF excimer laser ($\lambda$ = 248 nm). The laser energy density was in the range of 1-3 J/$cm^2$ and the repetition rate was 2 Hz. A ceramic disk of STO, processed by a standard ceramic synthesis technique, was used as a target. The depositions of the films took place at 750 ºC under the oxygen partial pressure of 50 mTorr. After the depositions, the films were cooled down to room temperature, keeping the oxygen pressure constant. The film thickness was varied in the range from 29 to 107 nm. The deposited STO films were epitaxially grown with (100) orientation. As an example, the x-ray diffraction (XRD) θ-2θ pattern of a 29 nm-thick STO film on LSAT substrate and the reflection high-energy electron diffraction (RHEED) image of this film along STO <110> are shown in Fig. 1. As it can be seen, a (100)-growth with highly crystallized epitaxial surface was obtained. Because of the lattice mismatch between STO (3.905Å) and LSAT (3.868Å), our STO films are in-plane compressed and out-of-plane elongated. In fact, the in-plane lattice parameter of the 49 nm thick film on LSAT substrate was 3.870±0.001 Å, which is very close to that of the substrate. On the other hand, the out-of-plane lattice parameter (3.936±0.001 Å) was substantially elongated. The out-of-plane lattice constant became slightly smaller with the increase of the thickness. Such out-of-plane elongation also occurred on NGO substrates [10] (3.942±0.001 Å for the 50 nm-thick film).

It has been known that the unit cell volume of STO films grown by physical vapor depositions such as pulsed laser deposition and sputtering is often larger than that of the bulk STO due to the oxygen deficiency, slightly off stoichiometry or other reasons [13-18]. Our previous study on STO films showed such a tendency [10], although this does not cause the considerable change in the basic strain feature, i.e. in-plane compressed and out-of-plane elongated, in the films on the used compressive substrates. The present films on LSAT substrates also indicated a similar tendency, which was however much less pronounced.

In addition to the pronounced reflection bands of the LSAT substrate, the IR reflection bands from the three polar phonons of cubic STO ($TO_1$, $TO_2$ and $TO_4$) are clearly seen in Fig. 2a, b at room temperature and 5 K, respectively. As expected, the strengths of modes increase with the film thickness and no other dramatic changes appear. The bare substrate reflectivity was measured for all the temperatures as a reference. The reflectivity spectra were fitted in a standard way using the generalized factorized damped harmonic oscillator model [19]. The fitted substrate parameters were fixed for fitting the two-slab (film + substrate) system. The dielectric function of the thin film was then evaluated using a classical model of independent oscillators [19]. The resulting dielectric function at selected temperatures for the 49 nm thin film is shown in Fig. 3a. The 107 nm film exhibits very similar complex dielectric function (not shown here) as the thinner one, only the peaks of ε' and ε'' near the $TO_1$ phonon frequency are higher in the thicker film because of the lower $TO_1$ phonon damping. Notice the low value of the static permittivity in Fig. 3a. It amounts to ~125 at room temperature and increases only up to ~230 at 5 K compared with the crystal value of ~300 at room temperature and more than 20,000 at 5 K. This is the consequence of huge in-plane soft mode stiffening in the strained film in comparison with the bulk samples. At room temperature the soft mode frequency amounts to 135 $cm^{-1}$ in the film and to ~90 $cm^{-1}$ in the crystal [20]. The soft mode frequencies at 5 K differ even more: 105 $cm^{-1}$ in the film compared to 15 $cm^{-1}$ in ceramics [4] or 8 + 17 $cm^{-1}$ (split doublet) in crystals [21].

Temperature dependences of the polar phonon frequencies as compared with the results from the STO/NGO system [10] are shown in Fig. 3b. The partial softening of the $TO_1$ soft mode is comparable to that in the STO/NGO system [10] and also the frequencies of the $TO_4$ mode and their temperature independence are comparable in both systems, stiffening from the bulk value of 542 $cm^{-1}$ to about 558 $cm^{-1}$. However, huge difference is observed in the case of $TO_2$ mode frequencies, which stiffens from the room-temperature bulk value of 173 $cm^{-1}$ to 183 $cm^{-1}$ in the STO/LSAT film and up to 200 $cm^{-1}$ in STO/NGO. The reason for such a pronounced difference is not understood. Another important difference between both types of film is that in STO/LSAT we do not observe the silent $TO_3$ mode, which in the case of STO/NGO film was interpreted as being activated by the ferroelectric phase transition near 150 K. Hence in our STO/LSAT films we do not observe any signature of the theoretically predicted out-of-plane ferroelectric transition. From the viewpoint of the strain state, both films are under the compressive in-plane strain, although the out-of-plane lattice constant of the film on LSAT is slightly smaller than that of the film on NGO. Such a similarity is basically well supported by the similar in-plane $TO_1$ soft mode frequency, which is known to be sensitive to the strain. On the other hand, the differences in $TO_2$ and $TO_3$ modes are still open questions. Further investigations like the impact of the in-plane asymmetric structure of (110) NGO substrate and the difference in the interfacial conditions is required.

We investigated also the polycrystalline STO film (40 nm thickness) deposited on the (0001) sapphire substrate at 650°C. The room-temperature x-ray (Cu K$\alpha$) diffraction studies showed a cubic perovskite structure with almost randomly oriented grains having a slight preferential (111) orientation. The lattice parameter $a$ determined from $\Theta$-2$\Theta$ reflections was $a$ = 3.905 Å, similar to that of the bulk crystal. Widths of the (110) and (111) peaks were corrected for instrumental broadening and analyzed using the Scherrer formula. The out-of-plane coherence length determined from the line widths was 35 – 40 nm, close to the film thickness, indicating that the grains extend through the film. AFM analysis of the surface morphology showed that the lateral size of such columnar grains was smaller than 100 nm.

The far-IR transmission spectra for selected temperatures up to 900 K of this film are shown in Fig. 4a. Because of the increasing absorption in the substrate, the soft mode parameters of the film can be well determined only below 300 K. A broad transmission minimum in the transmittance spectrum, corresponding to the $TO_1$ phonon in the STO film, is clearly seen at low temperatures (Fig. 4a). According to our knowledge, this is the first transmission measurement of the $TO_1$ soft mode in such a thin (40 nm) film. So far only substantially thicker films were investigated by the IR transmission [5]. The dielectric function obtained from the fits (evaluation procedure is described in Ref. [5]) is shown in Fig. 4b. At all temperatures, the soft mode appears overdamped, its response in the loss spectra being much broader than that in other STO films [5, 10], including STO/LSAT from this work. Considering that the film is relaxed and the effect of interfacial layers can be neglected since we probe the in-plane response, it appears that the grain boundaries and other film inhomogeneities which spread along the film plane could be the main reason for it, but the influence of point defects cannot be excluded. It is now well established that the grain boundaries play a pronounced role in the IR soft mode response in polycrystalline films as well as in bulk ceramics and their effect increases with the decreasing grain size [22, 23]. Note also the reduced static permittivity in the STO/sapphire 40 nm film, which increases only to ~400 at 20 K.

Concluding, we can state that the IR spectroscopy (both in transmission as well as reflection mode) is a useful contact-less tool for quantitative determination of the dielectric response of strong polar lattice vibrations down to the thicknesses of the order of several tens nm. Compressed STO films show appreciably stiffened polar-mode frequencies, particularly the ferroelectric soft mode, resulting in reduced static permittivity. In polycrystalline thin films the soft mode response is much broader than in epitaxial films or in bulk samples, probably due to the grain boundaries and other in-plane inhomogeneities. Thus the IR soft mode response appears to be very sensitive to the film quality, homogeneity and strains, which might be used for a standard contact-less characterization of the film quality.

The work was supported by the Grant Agency of Academy of Sciences of the Czech Republic (Projects KJB100100704 and AVOZ10100520), Ministry of Education (PC101-COST539) and by the Swiss National Science Foundation. NS and TM acknowledge the financial support of the EU (Nanostar project).

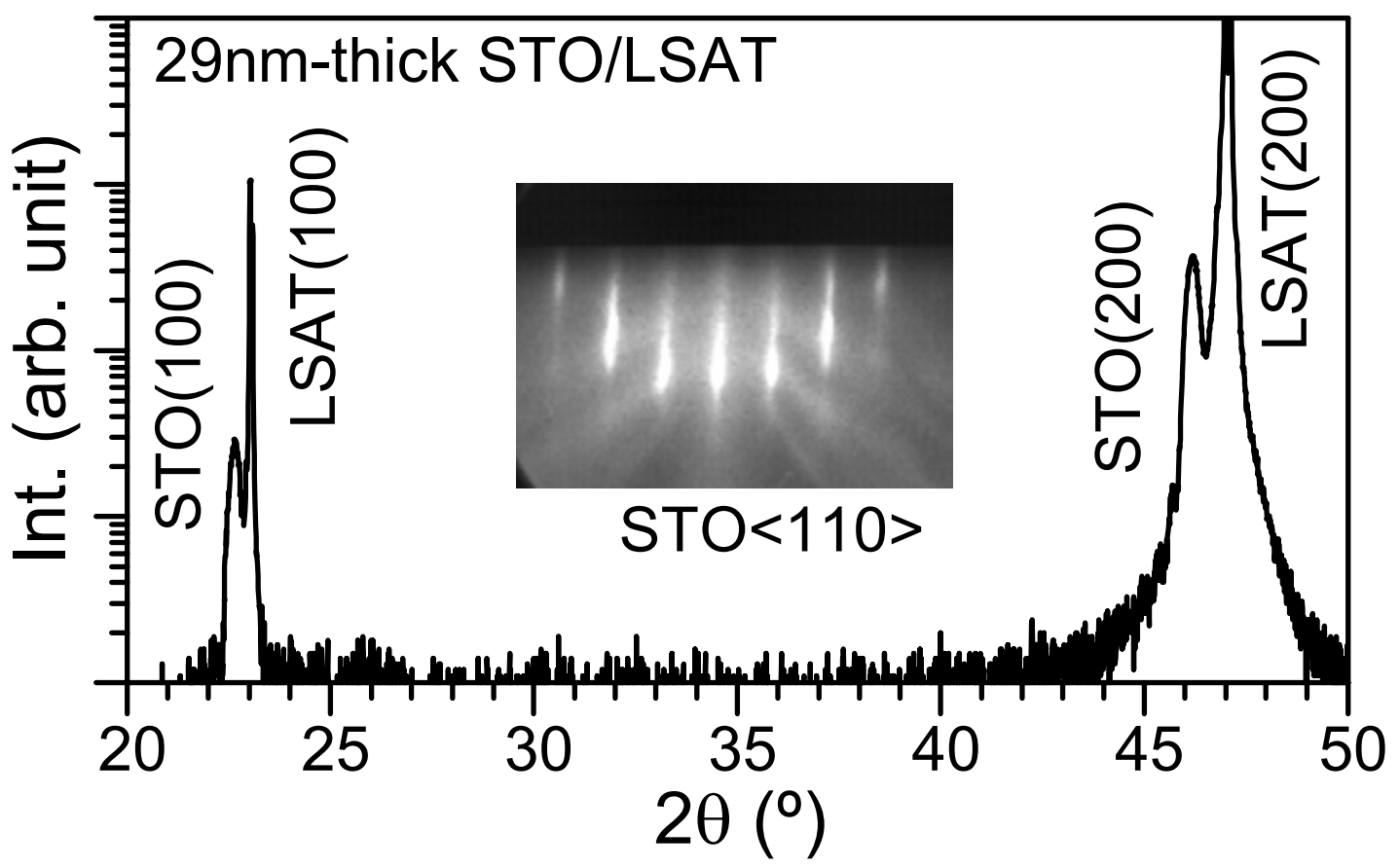

***Fig. 1.*** XRD θ-2θ scan and RHEED image of a 29 nm-thick STO film on the LSAT substrate. The incident electron beam axis is parallel to STO<110>.

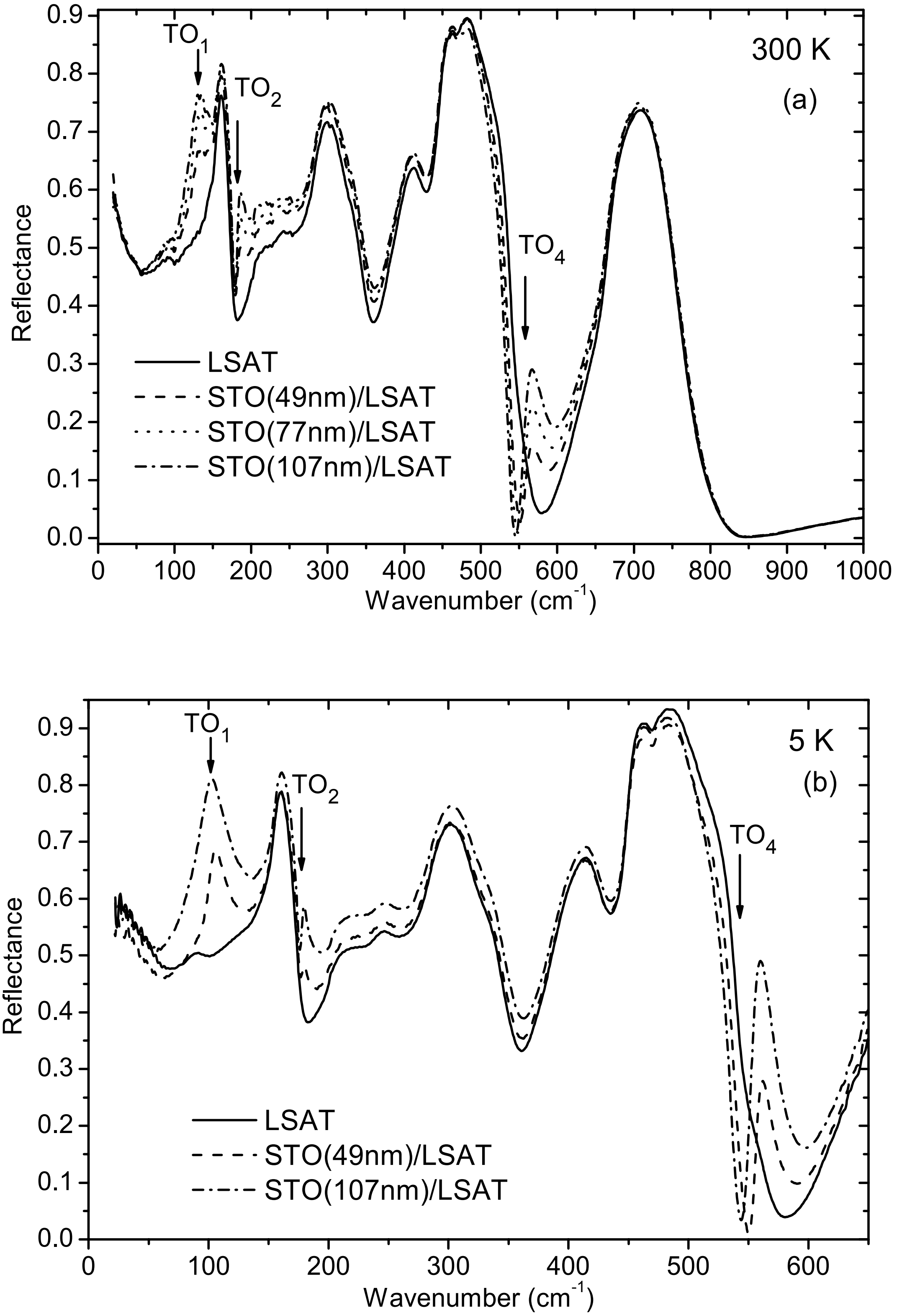

***Fig. 2.*** Infrared reflectance of the LSAT substrate and STO thin films of various thicknesses at (a) room temperature and (b) at 5 K.

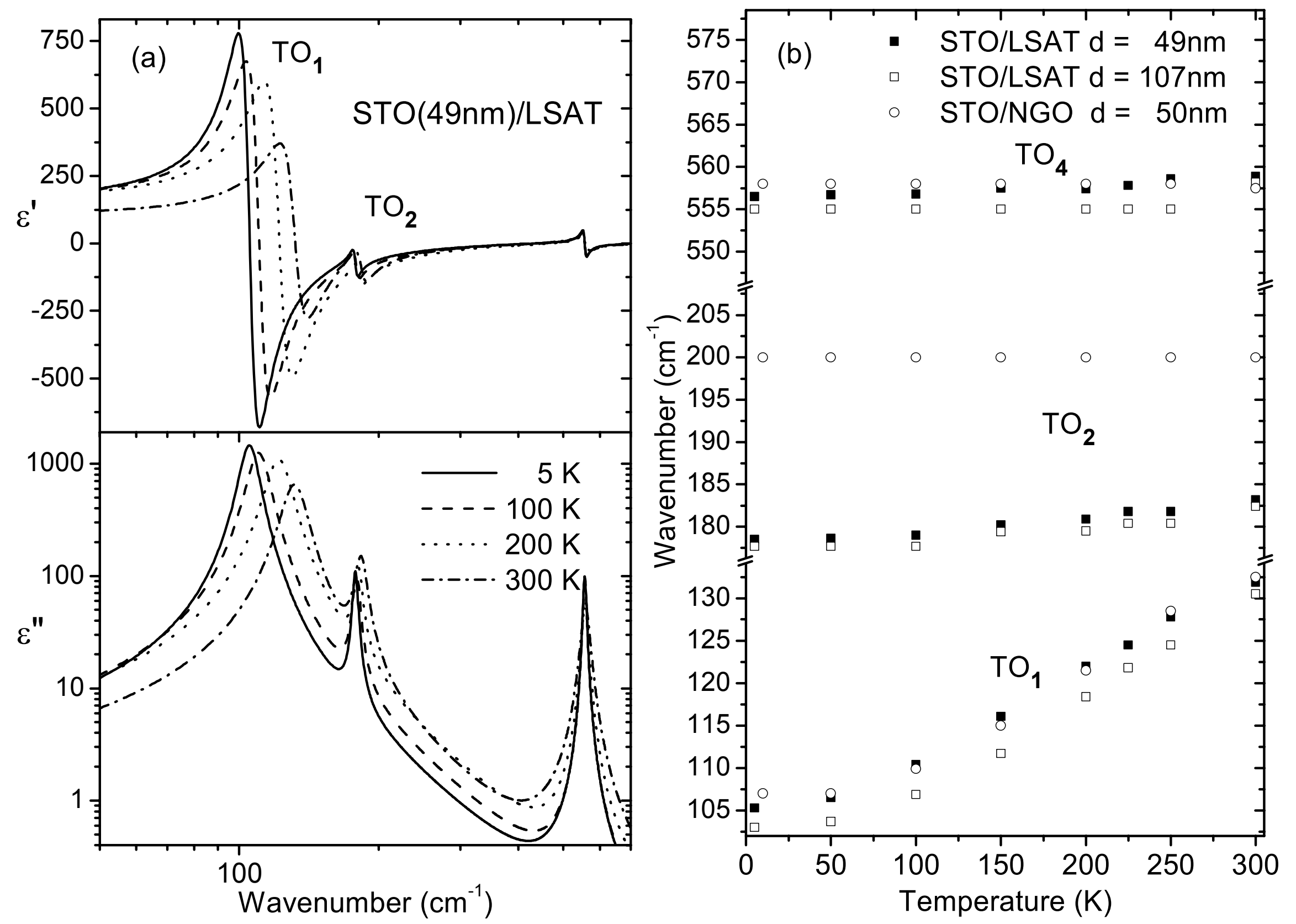

***Fig. 3.*** (a) Complex dielectric spectra of a 49 nm thick STO film deposited on LSAT substrate calculated from the fits to the far IR reflectance spectra. (b) Temperature dependence of the polar mode frequencies of various STO thin films on the LSAT and NGO substrates.

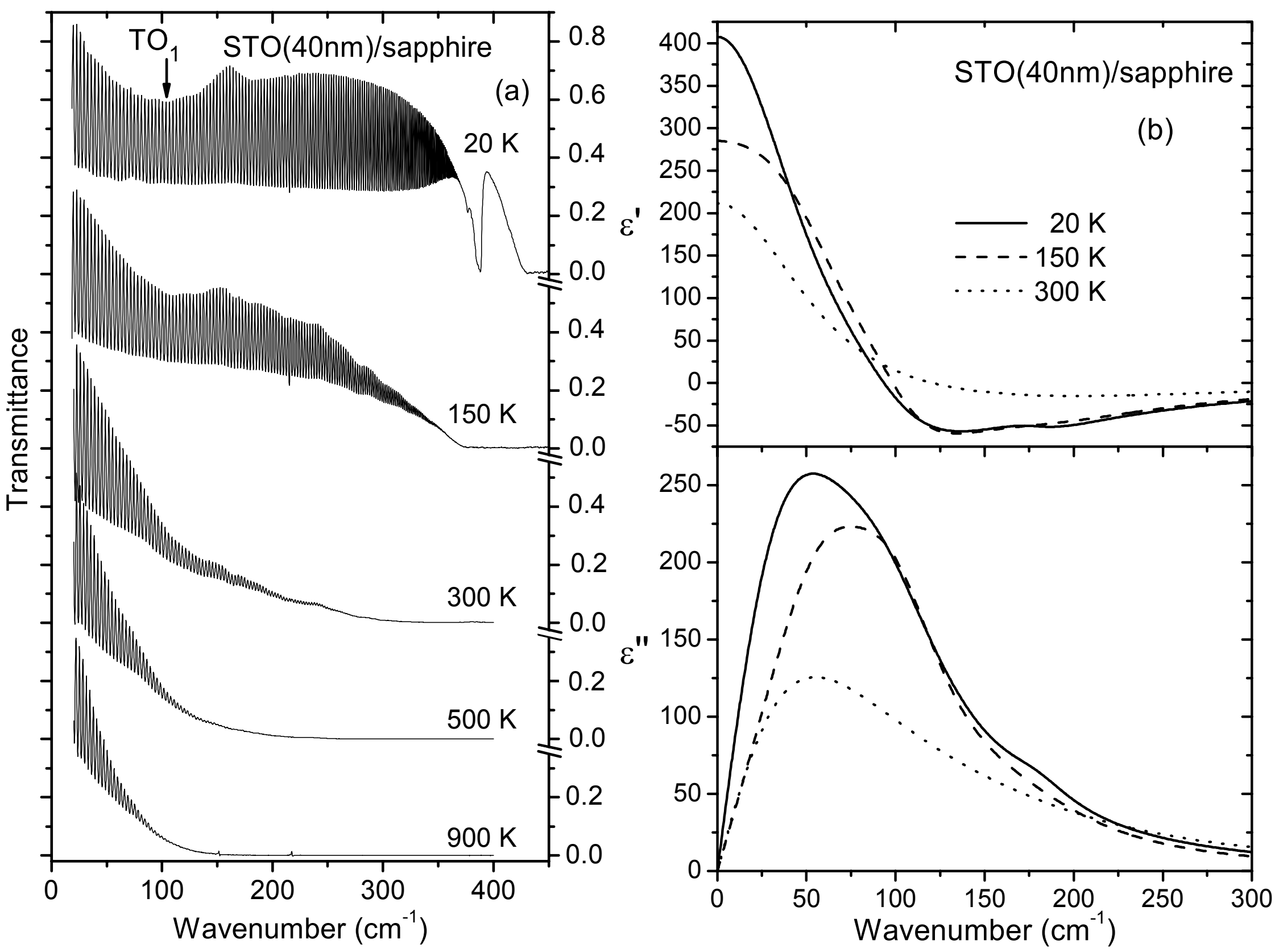

***Fig. 4.*** (a) Far IR transmittance of a 40 nm thick polycrystalline STO thin film deposited on the sapphire substrate. Dense oscillations in the spectra are interferences in the substrate, while the broad minima correspond to phonon frequencies of the thin film ($TO_1$ soft mode frequency is marked). (b) complex dielectric spectra of the STO film obtained from the fits to the far-IR transmittance spectra.